\documentclass[10pt]{iopart}

\usepackage{iopams}
\usepackage{graphicx}
\usepackage{bm}
\usepackage{enumitem}
\usepackage{xcolor}
\usepackage{tikz}
\usepackage{url}

\usepackage{todo}
\begin{document}

\title{Light scattering as a Poisson process and first-passage probability}
\author[1]{Claude Zeller}
\address{Claude Zeller Consulting LLC, Oceanside OR 97134}
\ead{czeller@ieee.org}
\author[2]{Robert Cordery}
\address{Fairfield University, 1073 North Benson Rd. Fairfield CT 06824}
\ead{rcordery@fairfield.edu}
\vspace{10pt}
\begin{indented}
\item[]June 2019
\end{indented}

\begin{abstract}
A particle entering a scattering and absorbing medium executes a random walk through a sequence of scattering events.
The particle ultimately either achieves first-passage, leaving the medium, or it is absorbed.
The Kubelka-Munk model describes a flux of such particles moving perpendicular to the surface of a plane-parallel medium with a scattering rate and an absorption rate.
The particle path alternates between the positive direction into the medium and the negative direction back towards the surface.
Backscattering events from the positive to the negative direction occur at local maxima or peaks, while backscattering from the negative to the positive direction occur at local minima or valleys.
The probability of a particle avoiding absorption as it follows its path decreases exponentially with the path-length \(\lambda\).
The reflectance of a semi-infinite slab is therefore the Laplace transform of the distribution of path-length that ends with a first-passage out of the medium.
In the case of a constant scattering rate the random walk is a Poisson process.
We verify our results with two iterative calculations, one using the properties of iterated convolution with a symmetric kernel and the other via direct calculation with an exponential step-length distribution.

We present a novel demonstration, based on fluctuation theory of sums of random variables, that the first-passage probability as a function of the number of peaks \(n\) in the alternating path is a step-length distribution-free combinatoric expression involving Catalan numbers.
Counting paths with backscattering on the real half-line results in the same Catalan number coefficients as Dyck paths on the whole numbers.
Including a separate forward-scattering Poisson process results in a combinatoric expression related to counting Motzkin paths. We therefore connect walks on the real line to discrete path combinatorics.
\end{abstract}
\section{Introduction}\label{sec:1Intro}
The perceived quality of a printed image is affected by three-dimensional light scattering in paper.
The Kubelka-Munk \cite{kubelka1931article} equations are most commonly used to estimate the effect of the medium on reflectivity.
Kubelka and Munk provide a one-dimensional analytic solution for a model with two fluxes on the positive half-line, one moving upward into the medium in the positive direction and the other moving in the negative direction, downward, back toward the interface at the origin.
While their approach is well-suited to analysis of reflectivity of layered media, it lacks the lateral scattering important to print quality.
A light ray entering a three-dimensional scattering medium executes a random walk through a series of scattering events and exits at a different point.
On a printed image this results in ``optical gain'' as light that would enter at a printed point and exit nearby is absorbed by the print.
The Yule-Nielsen~\cite{yule1951penetration} model extends Kubelka-Munk to describe optical gain.
While its physical basis is not completely established, the Yule-Nielsen is an improved print-quality model.
The impetus of this work is a more detailed understanding of scattering in one dimension as a step toward a true physically motivated three-dimensional scattering model with similar complexity to Kubelka-Munk.

The ubiquitous applications of random walks have inspired a large literature.
The monograph by Rudnick~\cite{rudnick2004elements} derives statistical properties of random walks and applies the theory to polymers and statistical mechanical systems.
Redner's book~\cite{redner2001guide} is devoted to first-passage processes.
Depending on the problem, these authors switch between continuous and lattice models.
The book by Spitzer~\cite{spitzer_principles_1964}, one of the pioneers in fluctuation theory, primarily addresses walks on a lattice.
Philips-Invernizzi and Caz\'{e}~\cite{philips-invernizzi_bibliographical_2001} provide a structured bibliographic review of the results obtained during the previous century.

Schwarzschild~\cite{schwarzschild1906equilibrium} investigated the temperature distribution within a stellar atmosphere using radiative equilibrium.
Chandrasekhar's  definitive theory for radiative transfer
\cite{chandrasekhar1950radiative, chandrasekhar1943stochastic} was developed in the context of plane-parallel layers of stellar and planetary atmospheres.
The solutions of radiation transport equations have a variety of applications from neutron diffusion, optical tomography~\cite{bonner1987model,nossal1988photon}, spreading of infra-red and visible light in the atmosphere to the quality of prints on paper.

The simplest solution of the radiative transfer equation is for a one-dimensional flux traveling perpendicular to a plane-parallel layer of an absorbing and scattering medium with isotropic radiation intensity over the forward and backward hemispheres.
One-dimensional radiative transfer can be solved by the two-flux approximation proposed independently by Schuster~\cite{schuster1905radiation}
and Schwarzschild~\cite{schwarzschild1906equilibrium} in astronomy.
Gate~\cite{gate1974comparison} examined the relation between Kubelka-Munk and the exact solution of the radiative transport equation.
More recently, Sandoval~\cite{sandoval_deriving_2014} generalized the Kubelka-Munk theory using radiative transfer theory.

Youngquist, Carr and Davies~\cite{youngquist1987optical} applied the model in optical coherence tomography and Haney and van Wijk~\cite{haney2007modified} in geology.
H\'{e}bert and Becker~\cite{hebert2008correspondence} examine the relation between continuous and lattice versions of Kubelka-Munk in an effort to give a more physical interpretation of the two-flux model.
Ballestra, Pacelli and Radi~\cite{ballestra2016very}  compute the first-passage probability density function using an integral representation obtained solving a system of convolution equations in the context of financial mathematics.

Simon and Trachsler~\cite{simon2003random} give an explicit expression for the reflectance.
They suggest that the scattering problem can be treated as a Markov chain involving Narayana polynomials.
But this Markov chain does not provide a solution of a first-passage problem that fits the reflectance calculated with the Kubelka-Munk equations.
Then using the compositional optical reflectance and transmittance properties for multilayer specimens, they determine a generating function and find a solution as variants of Chebyshev polynomials.
This is an alternative interpretation of the hyperbolic functions of the classical solution of Kubelka-Munk equations.

Monte Carlo simulations of random walks and first-passage process are commonly used to solve radiation transfer problems.
Even in one dimension, Monte Carlo simulations run into the problem that the mean number of scattering events before first-passage is infinite.
Jacques~\cite{jacques_monte_2010} developed code for scattering in multiple tissue layers.
Doering, Ray and Glasser~\cite{doering1992long} showed that the transmission time probability density has a long tail and anomalous moments due to multiple scattering events.
Antal and Redner~\cite{antal2006escape} study the first-passage of a random walk in an interval with a bounded uniformly distributed step length.
They observe non-diffusive effects that persist when the maximum step-length is small, especially if the starting point is close to one edge.

Wuttke~\cite{wuttke2014zig} recursively expands the equations of Darwin~\cite{darwin1922xcii} and Hamilton~\cite{hamilton1957effect}, or equivalently, the Kubelka-Munk equations.
He identifies Catalan numbers in the recurrence probability as a function of the number of backscattering events.
Wuttke defines a zigzag walk as an alternating walk where the random character comes from exponentially distributed step lengths between scattering events.
As we demonstrate, it is not necessary to specify the distribution since his result is independent of the distribution of step-length distribution.

Scattering with a constant rate is a Poisson process. It can be analyzed by solving differential equations, as illustrated in section~\ref{sec:2KM}.
Alternatively, the process can be treated as a random walk due to a sequence of scattering events with an exponentially distributed spacing.
The first-passage process is a probabilistic process by which a fluctuating quantity reaches a threshold for the first time.

One of our goals is to describe the Kubelka-Munk solution directly in terms of the statistics of the number of peaks \(n\) and the path-length \(\lambda\) of rays.
In one dimension the number of reflections before first-passage is \(2n-1\), so the scattering order, or number of reflections, is odd.

We explore Kubelka-Munk using a mathematically equivalent one-dimensional random walk model.
Rays entering the medium from above is a more natural perspective, but to make contact with the random walk language, we use rays entering upwards from the negative \(z\)-axis.
\begin{figure}
	\centering
	\tikzset{%
		dot/.style={circle, draw, fill=black, inner sep=0pt, minimum width=2pt},
		top/.style={anchor=south, inner sep=5pt},
	}
	\begin{tikzpicture}[xscale = 1, yscale=1,
	every label/.append style = {font=\footnotesize}]
	\draw [->] (0,0)-- (0,2) node (yaxis) [right] {$z$};
	\draw [->] (0,0)--(7,0) node (xaxis) [right] {step};
	
	\draw[dotted] (0, 0) -- (7, 0);
	\draw (0, 0) -- (1, 1.4) -- (2, 0.7) -- (3, 1.1) -- (4, 1.7) -- (5, .6) -- (6,-.5);
	\draw[double,->] (-.5,-1)--(0,0);
	\node[right] at (-.5,-1) {Incident flux};
	\node[dot] (P1) at (1, 1.4) {};
	\node[top] at (P1) {$z_1$};
	\node[dot] (P2) at (3, 1.1) {};
	\node[top] at (P2) {$z_3$};
	\node[dot] (P3) at (5, .6) {};
	\node[right] at (P3) {$z_5$};
	\node[dot] (V1) at (2, 0.7) {};
	\node[below] at (V1) {$z_2$};
	\node[dot] (V2) at (4, 1.7) {};
	\node[above] at (V2) {$z_4$};
	
	\node[dot] (E3) at (6, -.5) {};
	\node[below] at (E3) {$E_3$};
	\end{tikzpicture}
	\caption{The flux enters at step \(0\) traveling in the positive, or upward, direction. After several scattering events, the path leaves the medium. }
	\label{fig:Overview}
\end{figure}
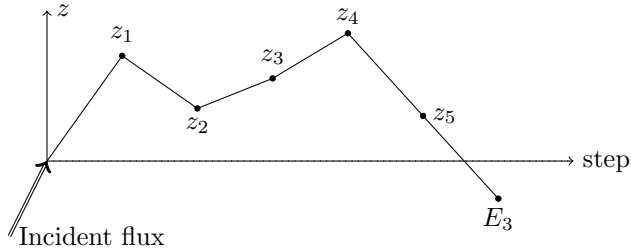
The scattering medium is on the positive \(z\)-axis.
The incident flux enters the medium moving upwards in the positive direction from negative \(z\).
An example path is illustrated in Fig.~\ref{fig:Overview}.
Backscattering or reflection occurs at \(z_1\), \(z_2\) and \(z_4\).
The step-lengths between scattering events are drawn from a random distribution.
We include an additional independent Poisson process for forward scattering with a rate \(S_f\).
The example path undergoes forward scattering at \(z_3\) and \(z_5\).
Usually forward scattering is not included in one-dimensional scattering because it does not change the ray propagation.
We include it here as a step towards future analysis of three-dimensional problems.
In three dimensions a scattering event may change the lateral direction, but not the direction perpendicular to the surface.
In this work, forward scattering and backscattering are assumed to be independent processes.
We demonstrate that the recurrence probability given by Catalan numbers is independent of the step-length distribution.
We provide another demonstration of the independence of step-length distribution using our formulation of the fluctuation theory introduced by Andersen~\cite{andersen1962equivalence, AndersenFluctuations}.

\section{Traditional solution of the Kubelka-Munk model}\label{sec:2KM}

This section illustrates the solution of the Kubelka-Munk~\cite{kubelka1931article} two-flux model as reviewed by Myrick, et al~\cite{myrick2011kubelka}.
Let us consider a homogenous layer with thickness \(d\) characterized by its absorption coefficient \(\chi\) and its scattering coefficient \(S\).
In this layer, the incident irradiance \(I\) propagates in the positive direction and the reflected irradiance \(J\) propagates in the negative direction.
Both \(I\) and \(J\) are functions of the depth \(z\) in the layer.
Depth 0 corresponds to the layer's boundary receiving the incident irradiance \(I_0\).
Depth \(d\) indicates the other boundary.
We consider, at an arbitrary depth \(z\), a sub-layer with infinitesimal thickness \(dz\).
The effect of the material in a thin element \(dz\) on \(I\) and \(J\) is to:
\begin{itemize}
  \item decrease \(I\) by \(I(S + \chi ) dz\)    (absorption and scattering)
  \item decrease \(J\) by \( J (S + \chi ) dz\)  (absorption and scattering)
  \item increase \(I\) by \(J S dz \)   (scattered light from \(J\) reinforces \(I\))
  \item increase \(J\) by \(I S dz \)   (scattered light from \(I\) reinforces \(J\)).
\end{itemize}

On these assumptions we obtain the system of equations:
\begin{equation}\label{eq2KMmatrix}
    \left( \begin{array}{c}
    {dI}\\
    {dJ}
    \end{array} \right) =
    \left( {\begin{array}{*{20}{c}}
    { - \left( {\chi  + S} \right)}&S\\
    { - S}&{\left( {\chi  + S} \right)}
    \end{array}} \right)
    \left( \begin{array}{c}
    I\\
    J
    \end{array} \right)dz,
\end{equation}
with solution
\begin{equation}\label{eq2KMgenSol}
    \left( \begin{array}{c}
    I(z)\\
    J(z)
    \end{array} \right) =
    \left( {\begin{array}{*{20}{c}}
    {1 - \beta }&{1 + \beta }\\
    {1 + \beta }&{1 - \beta }
    \end{array}} \right)
    \left( \begin{array}{c}
    A{e^{\kappa z}}\\
    B{e^{ - \kappa z}}
    \end{array} \right),
\end{equation}
where  \(\beta  = \sqrt {\chi\left(\chi  + 2S\right)^{-1}} \) and
\(\kappa  =  \sqrt {\chi \left( {\chi  + 2S} \right)}\).

The coefficients \(A\) and \(B\) are determined by the boundary conditions at the two surfaces.
After some elementary calculations reflectance \(R\) and transmittance \(T\) of a slab of thickness \(d\) are given by:
\begin{eqnarray}\label{eq2KMRT}
\eqalign{
    R & = \frac{J(0)}{I_0} = \frac{\left(1-\beta^2\right) \left(e^{\kappa d}- e^{-\kappa d}\right)}{\left(1+\beta\right)^2 e^{\kappa d}-\left(1-\beta\right)^2 e^{-\kappa d}} \\
    R_0 & = \frac{{Sd}}{{1 + Sd}}\quad for\;\chi  = 0 \\
    T &= \frac{{I(d)}}{{{I_0}}} = \frac{{4\beta }}{{{{\left( {1 + \beta }    \right)}^2}{e^{\kappa d}} - {{\left( {1 - \beta } \right)}^2}{e^{ -\kappa d}}}} \\
    T_0 &= \frac{{1}}{{1 + S d}}\textrm{ for }\chi  = 0 .
    }
\end{eqnarray}

The reflectance of a very thick layer \(\left(d\rightarrow\infty\right)\) is:
\begin{equation}\label{eq2KMRinf}
{R^{\infty}}(S,\chi ) = \frac{{S + \chi }}{S} - \sqrt {{{\left( {\frac{{S
+ \chi }}{S}} \right)}^2} - 1}.
\end{equation}
The Kubelka-Munk reflectance plays a major role in elucidating the connection to combinatorics and random walks.

\section{Distribution of path-lengths from the reflectance}\label{sec:3KMlength}
The fluxes \(I\) and \(J\) can be interpreted as ensembles of photons moving in the positive and negative directions.
Each photon path alternates at scattering events between positive and negative steps with a sequence of step lengths \( c_i \in \mathbf{c}\).
Step \(i\) has length \( c_i\) chosen from a probability distribution \(\rho \left( c_i \right)\) according to a Poisson process with rate \(S\) where
\begin{equation}\label{eq3rho}
    \rho \left( c \right) = \Phi (c)S{e^{ - Sc}}.
\end{equation}
\(\Phi\) is the Heaviside step function.
First, in the absence of absorption, the photon exits the medium after travelling a distance \(\lambda\), the path-length.
The probability distribution of \(\lambda\) for the random walk is \(P\left(\lambda;S\right)\).

Each photon is absorbed at a rate \(\chi\) per unit length.
Therefore, its path is weighted according to the Beer-Lambert law~\cite{beer1852bestimmung},~\cite{lambert1760photometria} by the absorption factor \(e^{-\chi\lambda}\).
The reflectance is therefore the Laplace transform of the path-length distribution and
the inverse Laplace transform of \(R^{\infty}\) leads to the path distribution:
\begin{eqnarray}
    {R^\infty }\left( {S,\chi } \right) &=& {{\mathcal L}_\lambda}\left\{ P\left(\lambda;S\right) \right\}\left(\chi\right) = \int\limits_0^\infty  {P\left( {\lambda ;S} \right){e^{ - \chi \lambda }}d\lambda }\\
\nonumber     P\left(\lambda;S\right) &=& {{\mathcal L}^{-1}_\chi}\left\{ R^\infty\left( {S,\chi } \right) \right\}\left(\lambda\right).
\end{eqnarray}

We calculate the inverse Laplace transform of the reflectance by first expanding equation~(\ref{eq2KMRinf}) in the scattering order
\begin{eqnarray}
  {R^{\infty}}(S,\chi ) &=& \sum\limits_{{n} = 1}^\infty {\frac{{{C_{{n} - 1}}}}{{{2^{2{n} - 1}}}}{{\left( {\frac{S}{{S + \chi }}} \right)}^{2{n} - 1}}} \\
\nonumber   &=& \frac{1}{2} \frac{S}{S + \chi }C\left(\left(\frac{1}{2} \frac{S}{S + \chi }\right)^2\right),
\end{eqnarray}\label{eq3KML-1}
where \(C(x)\) is the generating function
\begin{equation}\label{eqCatGen}
  C\left( x \right) = \sum\limits_{n = 0}^\infty  {{C_n}{x^n}}  = \frac{{1 - \sqrt {1 - 4x} }}{{2x}}
\end{equation}
of the Catalan numbers \({C_n} ={\left( {2n} \right)!}\left({{n!(n + 1)!}}\right)^{-1}\).
We identify, term-by-term, the path-length distribution of a random walk model with the inverse Laplace transform of \(R^{\infty}\left(S,\chi\right)\) using
\begin{equation}\label{eq3KML-1pterrms}
    {\mathcal{L}_{\chi}^{ - 1}} {{{\left( {\frac{S}{{S + \chi }}} \right)}^{2n - 1}}}  = \frac{{{S^{2n - 1}}{\lambda ^{2n - 2}}{e^{ - S\lambda }}}}{{(2n - 2)!}}.
\end{equation}

The distribution of \(\lambda\) and \(n\) derived from \(R^{\infty}\left(S,\chi\right)\)  is
\begin{eqnarray}\label{eq3KMPJ}
    P\left(\lambda ,n\right)&=
    \frac{1}{\lambda}{C_{n - 1}}\left(\frac{S\lambda}{2}\right)^{2n-1}
    \frac{e^{ - S\lambda}}{(2n - 2)!}\\
    &=\frac{1}{\lambda}\left(\frac{S\lambda}{2}\right)^{2n-1}
    \frac{{e^{ - S\lambda }}}{{n!(2n - 1)!}}.
\end{eqnarray}

The random character of an alternating walk comes from the step-length distribution between scattering events, which is not necessarily an exponential distribution.
We will demonstrate in the paper that certain statistics of alternating walks is independent of the step-length distribution, i.e., is distribution-free.

With no loss of generality we can include \(n_f\) forward scattering events at a rate \(S_f\) thus providing a persistent random walk model compatible with Kubelka-Munk.
Therefore the joint probability of the three random variables \(\lambda\), \(n\) and \(n_f\) is given by
\begin{equation}\label{eq3KMPJ2}
    P\left(\lambda ,{n},{n_f}\right) = \frac{{{C_{{n} - 1}}}}{{{2^{2{n} -
    1}}}}\frac{{{S^{2{n} - 1}}S_f^{{n_f}}{\lambda ^{2{n} - 2 +
    {n_f}}}}}{{(2{n} - 2)!{n_f}}}{e^{ - (S + {S_f})\lambda }},
\end{equation}
and the joint probability for \(n\) and \(n_f\) is then given by
\begin{eqnarray}\label{eq3KMPJG}
\eqalign{
    P\left({n},{n_f}\right) &= \int\limits_0^\infty  {P\left(\lambda,{n},{n_f}\right)} \,d\lambda\\
    &= \frac{1}{{{2^{2{n} - 1}}}}\frac{{\left[ {{n_f}
    + 2\left( {{n} - 1} \right)} \right]!}}{{{n_f}!{n}!\left( {{n} - 1}
    \right)!}}\frac{{S_f^{{n_f}}{S^{2{n} - 1}}}}{{{{\left( {{S_f} + S}
    \right)}^{{n_f} + 2{n} - 1}}}}.
    }
\end{eqnarray}

\begin{figure}
  \centering
      \tikzset{%
  dot/.style={circle, draw, fill=black, inner sep=0pt, minimum width=3pt},
  top/.style={anchor=south, inner sep=5pt},
}
    \begin{tikzpicture}[scale=.5]
        \draw [thin,->] (0,0)-- (0,3) node (yaxis) [right] {$z$};
        \draw [thin,->] (0,0)--(11.5,0) node (xaxis) [right] {step};
        \draw [thick] (0,0)--(1,1)--(2,1)--(3,2)--(4,2)--(5,1)--(6,0)--(7,1)--(8,1)--(9,1)--(10,0)--(11,0);
        \node[dot]  (0) at (0,0) {};
        \node[dot]  (1) at (1,1) {};
        \node[dot]  (2) at (2,1) {};
        \node[dot]  (3) at (3,2) {};
        \node[dot]  (4) at (4,2) {};
        \node[dot]  (5) at (5,1) {};
        \node[dot]  (6) at (6,0) {};
        \node[dot]  (7) at (7,1) {};
        \node[dot]  (8) at (8,1) {};
        \node[dot]  (9) at (9,1) {};
        \node[dot]  (8) at (10,0) {};
        \node[dot]  (9) at (11,0) {};
    \end{tikzpicture}
  \caption{A typical Motzkin path with eleven steps}\label{fig:motzkin}
\end{figure}
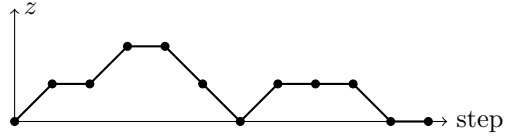

The combinatorial factor in equation~(\ref{eq3KMPJG}) is \(T\left(n_f+2 n -2, n-1\right)\) where
\begin{equation}\label{eq3Motz}
  T(n,k) = \frac{{n!}}{{k!\left( {k + 1} \right)!\left( {n - 2k} \right)!}}
\end{equation}
is the ``Triangular array of Motzkin polynomial coefficients.''
A Motzkin path is a path on an integer plane where each step is one of an up step \(\left(1,1\right)\), a level step \(\left(1,0\right)\) or a down step \(\left(1,-1\right)\).
According to the OEIS A055151~\cite{OEIS2019} \(T(n,k)\) is the number of Motzkin paths of length \(n\) with \(k\) up steps. The path shown in figure \ref{fig:motzkin} has length \(n=11\) with \(k=3\) up steps.

\section{Probability of first-passage by convolution}\label{sec:4FPconv}
\subsection{Iterations: building an alternating random walk}

The previous results are derived from the inverse Laplace transform of the reflectance.
We now want to study the trajectory distribution for a general path including both backscattering and forward scattering.
We assume that forward scattering and backscattering are independent processes.
We start with an alternating walk with only backscattering. The randomness comes only from the step length distribution.
The scattering event locations form an alternating sequence of maxima and minima (peaks and valleys).
First-passage necessarily occurs before a valley where the flux changes from the negative direction to the positive direction.
Therefore, we study first-passage in the alternating walk by marginalizing over peaks.
Finally, we introduce the forward scattering process.
The alternating walk serves as a skeleton of the general walk and we then “dress the skeleton” by adding forward scattering.

An alternating random walk, starts at 0 and moves upwards in the positive direction until it eventually backscatters downwards towards the origin.
The walk continues through a sequence of upwards motion, reflecting at a peak to the negative direction, and subsequent reflection at a valley back to the positive direction.
If a downwards step does not reverse direction before it reaches the negative half-space, we say it ``left the medium'' and the trajectory ends.
A trajectory that leaves the medium after \(n\) peaks is subject to \(\left( 2n-1\right)\) backscattering events and is labeled by the index \(n\).

Consider now the flux of all possible trajectories starting at \(z = 0\) and beginning in the positive direction with step length \(c\) distributed according to equation~(\ref{eq3rho}).

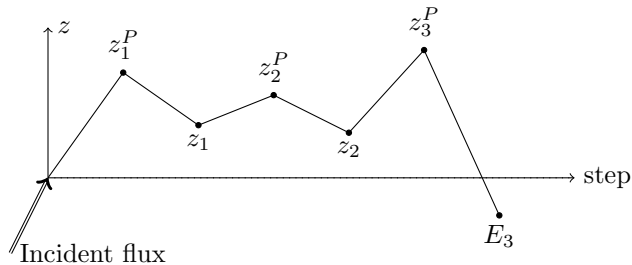
\begin{figure}
    \centering
    \tikzset{%
  		dot/.style={circle, draw, fill=black, inner sep=0pt, minimum width=2pt},
  		top/.style={anchor=south, inner sep=5pt},
	}
    \begin{tikzpicture}[xscale = 1, yscale=1,
    every label/.append style = {font=\footnotesize}]
    \draw [->] (0,0)-- (0,2) node (yaxis) [right] {$z$};
    \draw [->] (0,0)--(7,0) node (xaxis) [right] {step};
    \draw[dotted] (0, 0) -- (7, 0);
    \draw (0, 0) -- (1, 1.4) -- (2, 0.7) -- (3, 1.1) -- (4, .6) -- (5, 1.7) -- (6,-.5);
    \draw[double,->] (-.5,-1)--(0,0);
    \node[right] at (-.5,-1) {Incident flux};
    \node[dot] (P1) at (1, 1.4) {};
    \node[top] at (P1) {$z^P_1$};
    \node[dot] (P2) at (3, 1.1) {};
    \node[top] at (P2) {$z^P_2$};
    \node[dot] (P3) at (5, 1.7) {};
    \node[top] at (P3) {$z^P_3$};
    \node[dot] (V1) at (2, 0.7) {};
    \node[below] at (V1) {$z_1$};
    \node[dot] (V2) at (4, .6) {};
    \node[below] at (V2) {$z_2$};

  \node[dot] (E3) at (6, -.5) {};
  \node[below] at (E3) {$E_3$};
    \end{tikzpicture}
\caption{The flux enters at step \(0\) traveling in the positive, or upwards, direction. After several scattering events, the path can leave the medium. The figure shows an alternating path with first-passage at step six after the third peak. }
    \label{fig:samplePath}
\end{figure}
Marginalizing over the first peak height gives the probability distribution \({P_1}(z_1)\) of the height \(z_1\) of the first valley.
If \(z_1\) is negative, the trajectory ends at that point.
\({P_1}(z_1)\) serves as a convolution kernel for moving from one valley to the next with the following properties:
\begin{eqnarray}\label{eq4P1}
    \int_{ - \infty }^\infty  {{P_1}(z_1)dz_1}=1 \\
\nonumber    P_1(z_1) = P_1(-z_1).
\end{eqnarray}
\(P_1\) is symmetric in its argument because the upward and downward steps are drawn from the same distribution.

The trajectory reflects back towards the origin after the \(n^{\textrm{th}}\) peak at height \(z^P_n\) as shown in figure \ref{fig:samplePath}.
It reflects next time in the positive direction at the \(n^{\textrm{th}}\) valley at height \(z_n\).
The difference, \(z_n - z_{n-1}\), in height of the valleys is the difference of the upward and downward step lengths.
If \(z_n < 0\), then we say the trajectory left the medium.
The probability distribution for the height \(z_2\) of the valley at the second reflection is
\begin{eqnarray}\label{eq4Peaks}
    {P_2}\left( z_2 \right) = \int_{0}^{\infty} {{P_1}(z_1){P_1}(z_2-z_1)dz_1}\\
\nonumber    \int_{ - \infty }^\infty  {{P_{2}}(z_2)dz_2} = \int_0^\infty  {{P_{1}}(z_1)dz_1} =\frac{1}{2}.
\end{eqnarray}
This is a convolution of a symmetric function with itself.
The symmetry of \(P_1\) implies that, independent of the step-length distribution, half the trajectories leave the medium and half remain in the medium at the first valley.
This simple case suggests that first passage could be independent of the step-length distribution.

The probability distribution for the location of the following valleys \(n = 3, 4, \cdots\) are given by iterative convolutions:
\begin{equation}\label{eq4Pconv}
    {P_{n + 1}}\left( z_{n+1} \right) = \int_{0}^{\infty} {{P_{n}}(z_{n}){P_1}(z_{n+1}-z_{n})dz_{n}}.
\end{equation}
Only the trajectories that stay in the medium at step \(n\), are transferred to step \(n+1\), giving:
\begin{equation}\label{eq4Iterate}
    \int_{ - \infty }^\infty  {{P_{n + 1}}\left( z_{n+1} \right)dz_{n+1}}  =
    \int_0^\infty {{P_{n}}\left( z_{n} \right)} dz_{n}.
\end{equation}
The probability of not escaping the medium after \(n\) peaks is
\begin{equation}
\int_{-\infty}^{\infty}\int_0^{\infty}P_n\left(z_n\right)P_1\left(z_{n+1}-z_n\right)dz_n dz_{n+1} = P^{+}\left(n\right)
\end{equation}
and the probability of first-passage after \(n\) peaks is
\begin{equation}\label{Eq4:It_n}
\int_{-\infty}^{\infty}\int_{-\infty}^0 P_n\left(z_n\right)P_1\left(z_{n+1}-z_n\right)dz_n dz_{n+1} = P^f\left(n\right).
\end{equation}
with the property
\begin{eqnarray}
\sum_{m=1}^{n}\int_{-\infty}^{\infty}\int_{-\infty}^0 P_m\left(z_m\right)P_1\left(z_{m+1}-z_m\right)dz_m dz_{m+1} + \cdots\\
\nonumber \int_{-\infty}^{\infty}\int_0^{\infty}P_n\left(z_n\right)P_1\left(z_{n+1}-z_n\right)dz_n dz_{n+1}=1,
\end{eqnarray}
since the sum of the fractions that leave the medium after each peak plus the fraction that remain in the medium must add to 1.
Solving the iteration defined by equations~(\ref{eq4P1}) to~(\ref{Eq4:It_n}) requires an additional property.
Extensive Monte Carlo results with a variety of step-length distributions led us to believe that \(P^+(n)=(2n-1)P^f(n)\), independent of the choice of step-length distribution.
We solve the iteration inductively using this property giving results consistent with the Laplace transform of the reflectance in the previous section:
\begin{eqnarray}
P^{+}\left(n\right) &= \prod_{m=0}^{n-1} \frac{1+2m}{2+2m}=\frac{2n-1}{2^{2n-1}}C\left(n-1\right)\\
\nonumber 	P^f\left(n\right) &= \frac{1}{2^{2n-1}}C\left(n-1\right).
\end{eqnarray}

These results will be confirmed by an analytic calculation of the path distribution in the case of an exponentially distributed path-length, consistent with Kubelka-Munk, in section~\ref{sec:5Analytic}.  The case for a general path-length distribution is shown by a combinatorial method in section~\ref{sec:6DFpairing} equations~(\ref{eq6:pf}) and~(\ref{eq6:fpcalc}).

\subsection{Dressing the skeleton: connection with combinatorics}
The enumeration of lattice paths is a topic in combinatorics, closely related to the study of random walks in probability theory.
The ubiquitous presence of Catalan numbers in the joint distribution function
\(P\left({\lambda ,{n},{n_f};S,{S_f}} \right)\)
suggests a connection with combinatorics.
One approach to explain this connection is discretization.
Since the  statistics of first-passage is independent of the distribution of the path-length, to discretize the path we just have to integrate over \(\lambda\).
We expect this transition to a lattice model to reproduce the analytical result obtained in section \ref{sec:6DFpairing}.

The joint probability function is the product of the marginal probability times the conditional probability:
\begin{equation}\label{eq4PJF}
    P\left(n, n_f ; S, S_f\right)=P\left(n_f|n; S, S_f\right)P\left(n; S, S_f\right).
\end{equation}
The goal is to create new steps by randomly filling the \(2n\) segments of the skeleton with \(n_f\)  forward random scattering events and to calculate the resulting distribution.
Ultimately we want to find the related conditional probability \(P\left(n|n_f\right)\).
There are \(2n-1\) reflections and \({m_s} = 2{n} + {n_f}\) steps.

With our notation we have paths from \((2, 0)\) to \(2(n, n_f)\) with the constraint that \(n \geq 1\).
Therefore the number of paths \(N_C\) is given by:
\begin{equation}\label{eq4choosenpnf}
N_C =
\left( \begin{array}{c}
    2{n} + {n_f} - 2\\
    2{n} - 2
\end{array} \right).
\end{equation}
In terms of the number of scattering event \(m_s\) the number of paths is
\begin{equation}\label{eq4NoC}
    N_C({m_s},{n}) = \frac{{({m_s} - 2)!}}{{(2{n} - 2)!({m_s} -2{n})!}}.
\end{equation}

We can now write the conditional probability of \(m_s\) at constant \(n\) with \(r = \frac{S}{{S + {S_f}}}\)
\begin{equation}\label{eq4PC}
    P\left( {{m_s}|{n};r} \right) = \frac{{({m_s} - 2)!}}{{(2{n} - 2)!({m_s} - 2{n})!}}{r^{2{n} - 1}}{(1 - r)^{{m_s} - 2{n}}}.
\end{equation}

This result has been confirmed by extensive Monte Carlo calculations and is also obtained by recursion in section \ref{sec:5Analytic}.
Then the joint probability of \(n\) and \(n_f\) is given by
\begin{eqnarray}\label{eq4P_JG}
\eqalign{
   P\left( {{n},{n_f};r} \right) &= \frac{{\left[ {{n_f} + 2\left( {{n} - 1} \right)} \right]!}}{{{n_f}!{n}!({n} - 1)!}}{\left( {\frac{r}{2}}
    \right)^{2{n} - 1}}{\left( {1 - r} \right)^{{n_f}}}\\
    &= \frac{{\left( {{n_f} + 2\left( {{n} - 1} \right)} \right)!}}{{{2^{2{n}
    - 1}}{n_f}!\left( {{n} - 1} \right)!({n})!}}\frac{{{S^{2{n} -
    1}}{{\left( {{S_f}} \right)}^{{n_f}}}}}{{{{\left( {S + {S_f} + \chi }
    \right)}^{{n_f} + 2{n} - 1}}}}.
    }
\end{eqnarray}
This is the same as equation~(\ref{eq3KMPJG}).
This confirms the combinatorial nature (lattice paths enumeration) of the discrete form of the Kubelka-Munk equation.

\section{Analytic calculation of the path distribution}\label{sec:5Analytic}

In this section, explicit integration over the path gives the distribution of path-length and number of scattering events as a function of depth into the medium.
We include forward scattering, reflection and absorption as independent processes acting on an ensemble of light rays.
We show directly the connection to equation~(\ref{eq2KMRinf}), as opposed to the two-step calculation in section~\ref{sec:4FPconv}.
We derive information about the distribution of the number of scattering events and the path-length when the ray ultimately escapes.
The current state of a ray at a scattering event includes the direction, the total prior path-length \(\lambda\), the prior number of peaks \(n\) and the position \(z\).
This will serve as a foundation for future exploration of higher dimensions and for the effect of inhomogeneities in the medium that are important for print quality and medical imaging.

Consider a statistical ensemble of rays of light moving in a one-dimensional diffusive medium on the positive axis.
Again, the ray starts at the origin moving upward in the positive direction and after multiple reflections, escapes the media to the negative axis.
A ray traveling through the medium is reflected at a rate of \(S\) reflections per unit length.
The Poisson probability density for reflection of a ray after traveling distance \(c\) is given by equation~(\ref{eq3rho}).
Later we will add forward scattering at a rate \(S_f\) and absorption at a rate \(\chi\) per unit length as independent Poisson processes based on the path-length.

The probability density for an upward reflection at a valley at height \(z\) with current path-length \(\lambda\) at the valley after \(n\) peaks is  \( {P_n}\left( {\lambda,z} \right) \).
The probability density that the ray escapes with total path-length \(\lambda\) in the scattering medium, after the  \(n^\textrm{th}\) peak is \(E_{n}\left(\lambda\right)\).
The probability densities for the peaks and valleys are not normalized because the ray may escape on a previous step.

The initial path-length is 0 and the initial position is the origin.
The first step is into the medium, so the initial escape probability \(E_0 = 0\).
\begin{eqnarray}\label{EqInitCond}
  P_0\left(\lambda,z\right) &=& \delta\left(\lambda\right)\delta\left(z\right) \\
 \nonumber E_0\left(\lambda\right) &=& 0.
\end{eqnarray}

The first peak integral simply involves satisfying the delta functions.
Similarly, the first valley integral uses the fact that \(\lambda\) is the total path-length.
The joint probability distribution of the height and path-length for the first peak is
\begin{eqnarray}\label{EqFirstPeak}
\fl P_1^P\left( {\lambda ,{z^P}} \right) &=& \int_0^\lambda {\int_0^\infty  {{P_0}\left( {\lambda',z'} \right)\delta \left( {\lambda  - \left( {\lambda' + z^P - z'} \right)} \right)\rho \left( {z^P - z'} \right)d\lambda'dz'} } \\
\fl \nonumber  &=& \delta \left( {\lambda -{z^P}} \right)\rho \left( {{z^P}} \right).
\end{eqnarray}
Integrating over the first peak gives the joint distribution of the path-length \(\lambda\) and height \(z_1\) of the first valley
\begin{eqnarray}\label{EqFirstValley}
 P_1\left( {\lambda ,{z_1}} \right) &=& \int_0^{\infty}dz_1^P
                 \delta \left( {\lambda  - {2z_1^P}+z_1} \right)
S{e^{ - S z_1^P }} \rho\left(z_1^P-z_1\right)   \\
 \nonumber &=& \frac{1}{2} \Phi(z_1)\Phi(\lambda) \Phi(\lambda-z_1) S^2e^{-S\lambda}.
\end{eqnarray}

The probability that the ray escapes is given by dropping the constraint that \(z\) is positive and integrating over the negative half space:
\begin{eqnarray}\label{eqEsc1lambda}
    {E_1}\left( \lambda \right)
     =  \frac{1}{2}\Phi \left( \lambda \right)
    \int_{ - \infty }^0
    {{S^2}{e^{ -S\left( {\lambda - z} \right)}}dz}
     = \frac{1}{2}\Phi \left(\lambda\right)
    S{e^{ - S\lambda}}.
\end{eqnarray}
Integrating over \(\lambda\) gives the expected escape rate of \(\frac{1}{2}\) after the first peak.

The probability distribution of the \(n^\textrm{th}\) peak and valley are found by iteratively calculating the convolution using an upward step with length \(c'\) and a downward step with length \(c\) giving
\begin{eqnarray}
\fl   P_n^P\left(\lambda,z_n^P\right) &=& \int_0^{{z_n^P}} {P_{n - 1}\left( {\lambda - c',z_n^P-c'} \right)S{e^{ - Sc'}}dc'} \\
\fl \nonumber  P_n\left(\lambda,z_n\right) &=& \int_z^\infty {P_{n}^P\left( {\lambda - c,z_n + c} \right)S{e^{ - Sc}}dc}\\
\fl \nonumber P_n\left( {\lambda ,{z_n }} \right) &=& \int_0^\infty  {\int_0^{c + {z_n}} {P_{n - 1}\left( {\lambda  - c' - c,{z_n} + c - c'} \right){S^2}{e^{ - S\left( {c + c'} \right)}}dc'} dc}.
\end{eqnarray}

There are constraints on the integrals in iterating from one valley to the next.
The peaks are higher than the preceding and following valleys, i.e., \(z_n^P \ge z_n\) and \(z_n^P \ge z_{n - 1}\), as shown in figure~\ref{fig:samplePath}.

Iterating from the probability distribution at one valley to that at the next valley, we obtain:
\begin{eqnarray}
P_n\left( \lambda,z_n \right)  =  S^{2n}e^{ - S\left( \lambda \right)}
\Phi \left( {\lambda - z_n} \right){F_n}\left( {\lambda,z_n} \right).
\end{eqnarray}
Here \(F_{n}\) is the total volume of the space of allowed configurations of the \(2n-2\) step path from the origin to a valley at the point \(z\) with path-length \(\lambda\).
The same form, an exponential times the volume of path configurations, will apply in higher dimensions and more complicated geometries.
Monte Carlo methods can be applied to measure the path configuration volume in these situations.

\begin{eqnarray}
{F_{{n}}}\left( {\lambda, z} \right) &=&
\frac{\left(\lambda-z\right)^{n-1}\left(\lambda+z\right)^{n-2}\left(\lambda-z+2n z\right)^{n-1}}{2^{2n-2}\left(n-1\right)!\left(n\right)!}.
\end{eqnarray}

Integrating over the negative half-space gives the escape probability density as a function of upward path-length and the total escape probability after the  \(n^{th}\) peak:
\begin{eqnarray}
\eqalign{
\fl
{E_{{n}}}\left( \lambda \right) &= \int_{ - \infty }^0 {{P_{{n}}}\left( {\frac{\lambda}{2},z'}\right)dz'}
= S{\left( {S\frac{\lambda}{2}} \right)^{2\left( {{n} - 1}
\right)}}\frac{{{e^{ - S\lambda}}}}{\left( {{n} - 1} \right)!\left(n \right)!}.
}
\end{eqnarray}

Integrating over \(\lambda\) gives the escape probability after \(n\) peaks in terms of Catalan numbers.
The result is independent of the scattering rate and the formula is identical to equation~(\ref{eq3KML-1}) with \(\chi = 0\):
\begin{equation}
{E_{{n}}} = \int_0^\infty  {{E_{{n}}}\left( \lambda \right)d\lambda} =
\frac{1}{{{2^{2{n} - 1}}}}{C_{{n} - 1}}.
\end{equation}

We can add in absorption because we have the distribution of the path-length.
The attenuation simply adds \(\chi\) to \(S\) in the exponent:
\begin{equation}
    E_{{n}}^\chi\left( \lambda \right) =
    S{\left( {S\lambda} \right)^{2{n} - 1}}\frac{{{e^{ - \left( {S + \chi } \right)\lambda}}}}{{\left( {{n} - 1}\right)!\left( {{n}} \right)!}}.
\end{equation}
Integrating over the path-length again yields equation~(\ref{eq3KML-1})
\begin{eqnarray}
\eqalign{
E_{{n}}^\chi = \int_0^\infty  {E_{{n}}^\chi (\lambda)d\lambda = {{\left(
{\frac{S}{{S + \chi }}} \right)}^{2{n} - 1}}\frac{{\left( {2{n} - 1}
\right)!}}{{{2^{2{n} - 1}}\left( {{n} - 1} \right)!\left( {{n}}
\right)!}}} \\
\quad \quad  = {\left( {\frac{S}{{S + \chi }}} \right)^{2{n} -
1}}\frac{{{C_{{n} - 1}}}}{{{2^{2{n} - 1}}}}.
}
\end{eqnarray}

The joint probability for \(n\) and \(n_f\) is obtained by adding an independent Poisson process for forward scattering
\begin{eqnarray}
\eqalign{
    P\left({n_f},{n}\right) &= \int_0^\infty
    {P_{\textrm{Pois}}({n_f}|{S_f}\lambda)E_{{n}}^\chi (\lambda)d\lambda}\\
    &= \frac{{\left( {{n_f} + 2\left( {{n} - 1} \right)} \right)!}}{{{2^{2{n}
    - 1}}{n_f}!\left( {{n} - 1} \right)!({n})!}}\frac{{{S^{2{n} -
    1}}{{\left( {{S_f}} \right)}^{{n_f}}}}}{{{{\left( {S + {S_f} + \chi }
    \right)}^{{n_f} + 2{n} - 1}}}}.
    }
\end{eqnarray}
This result is identical to equations~(\ref{eq3KMPJG} and~\ref{eq4P_JG}).

\section{First-passage events are distribution-free}\label{sec:6DFpairing}

This section extends the fluctuation theory of Andersen~\cite{AndersenFluctuations} to alternating walks with independent and identically distributed step-lengths.
We prove first-passage is distribution-free by examining first-passage properties of the finite set of alternating walks generated from permutations of a set of \(2n\) step-lengths.
Surprisingly, the probability of first-passage at each valley for this set of walks is independent of the set of step-lengths.

Consider the sample space \(\mathcal{A}_n\) of alternating walks with \(2n\) steps.
In this section we use the capitalized term ``Event'' in the probabilistic sense.
An Event is a subset of the sample space of \(\mathcal{A}_n\).
The indicator function \(\mathbb{I}_E\) of an Event is \(1\) for a walk that is an element of the Event and \(0\) otherwise.
The probability of an Event is given by the average over \(\mathcal{A}_n\) of the indicator function \(\mathbb{I}_E\) of the Event.
First-passage Events are the subsets \(\mathcal{F}_m\) of \(\mathcal{A}_n\) that are positive before peak \(m\) and first become negative at step \(2m\) after peak \(m\).
It is possible that a walk never becomes negative, so we add the Event \(\mathcal{F}_0\) of walks that are positive for the first \(2n\) steps.
The set of \(n+1\) first-passage Events, one at each valley plus one if the walk stays positive, is a partition of the sample space.

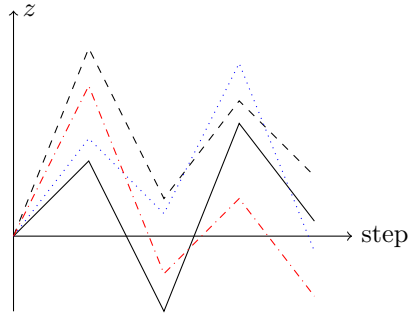
\begin{figure}
    \centering
    \tikzset{%
  		dot/.style={circle, draw, fill=black, inner sep=0pt, minimum width=2pt},
  		top/.style={anchor=south, inner sep=5pt},
	}
    \begin{tikzpicture}[xscale = 1, yscale=1,
    every label/.append style = {font=\footnotesize}]
    \draw [->] (0,-1)-- (0,3) node (yaxis) [right] {$z$};
    \draw [->] (0,0)--(4.5,0) node (xaxis) [right] {step};
    \draw (0, 0) -- (1, 1) -- (2, -1) -- (3, 1.5) -- (4, .2);
    \draw [dashed] (0, 0) -- (1, 2.5) -- (2, .5) -- (3, 1.8) -- (4, .8);
    \draw [dotted, blue] (0, 0) -- (1, 1.3) -- (2, .3) -- (3, 2.3) -- (4, -.2);
    \draw [dashdotted, red] (0, 0) -- (1, 2) -- (2, -.5) -- (3, .5) -- (4, -.8);
    \end{tikzpicture}
\caption{Four examples of the 24 alternating walks generated from permutations of an example set of four lengths  \(\mathbf{c}_2 =\{1.0, 1.3, 2.0, 2.5\}\). }
    \label{fig:stepPermutation}
\end{figure}

The location of an alternating walk alternates between peaks and valleys as shown in figure~\ref{fig:samplePath}.
First-passage for an alternating walk with \(n\) peaks occurs at a valley where the step number \(2m\) is even.
The positions of the peaks and valleys are given by repeated convolution with the step-length distribution as illustrated in sections~\ref{sec:4FPconv} and~\ref{sec:5Analytic}.
The set of \(2n\) step-lengths \(\mathbf{c}_n\) inside the integrand of the iterated convolution is an element of the set \(\mathcal{C}_n\) of all sets of \(2n\) lengths.
Because the step-lengths are independent and identically distributed, the joint probability density in the integrand is symmetric under permutations of the \(2n\) step-lengths \(\mathbf{c}_n\) in the integrand.
The average of the indicator function \(\mathbb{I}_E\) over \(\mathcal{A}_n\) can therefore be evaluated by first averaging over the finite set of walks generated by permutations of \(\mathbf{c}_n\).
Figure~\ref{fig:stepPermutation} illustrates, for example, four of the 24 walks generated from permutations of a particular  \(\mathbf{c}_2\).
The surprising result, shown below, is that this average over a finite set of walks is independent of the set \(\mathbf{c}_n\).

Consider the finite set of walks generated by permutations of \(\mathbf{c}_n\).
Each of the alternating walks in \(\mathcal{A}_n\left( \mathbf{c}_n\right)\) generated by permutations of \(\mathbf{c}_n\) has the same weight in the integral.
We will show that the cardinality of \(\mathcal{F}_m\left( \mathbf{c}_n\right)\) is independent of the set of \(2n\) step lengths \(\mathbf{c}_n\) almost everywhere, as long as no sub-walk returns exactly to the origin.

We analyze the changes in Event membership as we modify step-lengths in \(\mathbf{c}_n\).
The boundary subset \(\mathcal{B}_n\) of \(\mathcal{C}_n\) of measure zero, consists of those sets of step-lengths \(\mathbf{c}_n\) where some sub-walk constructed from \(\mathbf{c}_n\) returns to the origin.
For a walk constructed from a set of lengths not in the boundary set \(\mathcal{B}_n\), let \(\delta\) be the closest approach to the origin of any non-empty sub-walk.
At step \(2m > 0\) the absolute value of the position must be at least \(\delta\).
If \(\delta > 0\) then any step-length change with magnitude smaller than \(\delta\) will not cause any change in first-passage Event membership.
First-passage Event cardinality is therefore locally constant, and so is constant within connected subsets of \(\mathcal{C}_n - \mathcal{B}_n\).
The set \(\mathcal{C}_n - \mathcal{B}_n\) is not connected, so we must examine changes in Event cardinality when crossing a boundary.

Any set of lengths can be reached by changing one length at a time, and so if cardinality of an Event never changes when crossing the boundary \(\mathcal{B}_n\), then the Event cardinality is constant in \(\mathcal{C}_n - \mathcal{B}_n\).
The sequences of lengths in \(\mathcal{B}_n\) are the only elements of \(\mathcal{C}_n\) where walks can leave or enter first-passage Events with small changes to a length.
Consider changing a length in a walk \(w_m\) in \(\mathcal{F}_m\) with first-passage following the \(m\)th peak. It can move across the boundary set \(\mathcal{B}_n\).  As it crosses the boundary point the membership of \(w_m\) in  \(\mathcal{F}_m\) can change.
The set of walks generated from permutations or an overall sign change of the steps in the sub-walk \(w^*_m\) also return to the origin.
The problem is to show cancellation of the changes in Event membership among the walks in this set as lengths are changed.

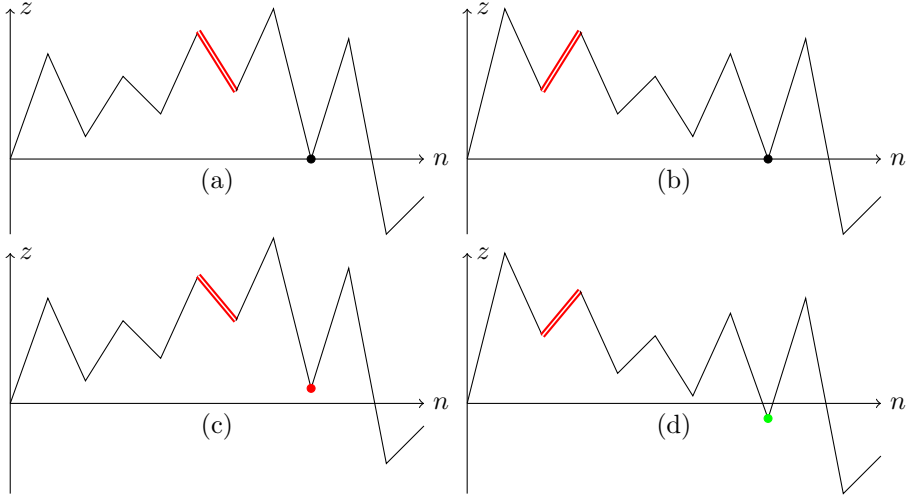
\begin{figure}
      \tikzset{%
      dot/.style={circle, draw, fill=black, inner sep=0pt, minimum width=3pt},
      top/.style={anchor=south, inner sep=5pt},
    }
    \centering
    \begin{tikzpicture}[xscale = 0.5, yscale=1]
        \draw[->] (0, 0) --  node[below] {(a)} (11, 0) node (xaxis) [right] {$n$};
        \draw[->] (0, -1) --  (0,2) node (yaxis) [right] {$z$};
        \draw (0, 0) -- (1, 1.4) -- (2, 0.3) -- (3, 1.1) -- (4, .6) -- (5, 1.7);
        \draw[thick, double, red] (5, 1.7) -- (6, 0.9);
        \draw (6, 0.9) -- (7, 2) -- (8, 0) -- (9, 1.6) -- (10, -1) -- (11, -.5);
        \node[dot]  (1) at (8,0) {};
    \end{tikzpicture}
    \centering
	\begin{tikzpicture}[xscale = 0.5, yscale=1]
        \draw[->] (0, 0) --  node[below] {(b)} (11, 0) node (xaxis) [right] {$n$};
        \draw[->] (0, -1) --  (0,2) node (yaxis) [right] {$z$};
        \draw (0, 0) -- (1, 2) -- (2, 0.9);
        \draw[thick,  double, red] (2, 0.9)--(3, 1.7);
        \draw (3, 1.7) -- (4, .6) -- (5, 1.1) -- (6, 0.3) -- (7, 1.4) -- (8, 0) -- (9, 1.6) -- (10, -1) -- (11, -.5);
        \node[dot]  (1) at (8,0) {};
    \end{tikzpicture}
    \centering
    \begin{tikzpicture}[xscale = 0.5, yscale=1]
    \draw[->] (0, 0) --  node[below] {(c)} (11, 0) node (xaxis) [right] {$n$};
    \draw[->] (0, -1.2) --  (0,2) node (yaxis) [right] {$z$};
    \draw (0, 0) -- (1, 1.4) -- (2, 0.3) -- (3, 1.1) -- (4, .6) -- (5, 1.7);
    \draw[thick, double, red] (5, 1.7) -- (6, 1.1);
    \draw (6, 1.1) -- (7, 2.2) -- (8, 0.2) -- (9, 1.8) -- (10, -.8) -- (11, -.3);
        \node[dot,red]  (1) at (8,0.2) {};
    \end{tikzpicture}
    \centering
	\begin{tikzpicture}[xscale = 0.5, yscale=1]
    \draw[->] (0, 0) --  node[below] {(d)} (11, 0) node (xaxis) [right] {$n$};
    \draw[->] (0, -1.2) --  (0,2) node (yaxis) [right] {$z$};
    \draw (0, 0) -- (1, 2) -- (2, 0.9);
    \draw[thick, double, red] (2, 0.9)--(3, 1.5);
    \draw (3, 1.5) -- (4, .4) -- (5, .9) -- (6, 0.1) -- (7, 1.2) -- (8, -.2) -- (9, 1.4) -- (10, -1.2) -- (11, -.7);
        \node[dot,green]  (1) at (8,-.2) {};
    \end{tikzpicture}
    \caption{Example of pairing critical walks for the Event ``First-passage on step 8.'' Critical walk (a) is paired with critical walk (b) with the sub-walk for the first eight steps reversed. When the sixth step length is shortened, (c) shows walk (a) leaving the Event and (d) shows walk (b) joining the Event. }
    \label{fig:2criticalWalk}
\end{figure}

The key step in our approach is to consider uniquely defined pairs of walks \(w^*\) and \(w'^*\) from the boundary set.
Walk \(w^*\) begins with a positive critical sub-walk \({w^*_{m}}\) as in figure \ref{fig:2criticalWalk}(a), but returns precisely to zero at step \(2m\) as illustrated by the dot at step 8.
The paired walk \(w'^{*}\) in figure \ref{fig:2criticalWalk}(b) is identical after step \(2m\) but begins with the time-reversed sub-walk \(w'^{*}_m\).
The sets of positions of the paired walks are the same, but they are in the reverse order for the first \(2m\) steps.
The step lengths in \(w'^*_m\) are the same as the step lengths in \(w^*_m\), but the first \(2m\) steps occur in the reverse order and with the opposite sign.
All the critical walks generated from \(\mathbf{c^*}_n \in \mathcal{B}_n\) that could change the cardinality of \(\mathcal{F}_m\) can be uniquely paired in this way.

For the first-passage problem, \(\mathcal{F}_m\) consists of walks that are positive for the first \(2m-1\) steps and negative on step \(2m\). The boundary of \(\mathcal{F}_m\) consists of walks that are non-negative for the first \(2m-1\) steps but return exactly to zero at one of the first \(2m\) steps.
Walks that return precisely to the origin on step \(2m\) and are positive before that step for steps \(1, \cdots \left(2m-1\right)\) will enter or leave the Event with a small change in a length.
Similarly, walks that pass zero on step \(2m\) but return precisely to the origin on step \(2j < 2m\) change membership in \(\mathcal{F}_m\) with small length changes.

Suppose a walk from the boundary of \(\mathcal{F}_m\) begins with such a \(2j\)-step sub-walk \({w^*_{j}}\) which is positive for \(k < 2j\) and returns to \(0\) at step \(2j\).
Consider changing a length \(c_k \in w^*_j\) through the critical value \(c^*_k\).
A small decrease in a length that is a downward step in \({w^*_j}\) will cause the walk to leave \(\mathcal{F}_j\) and join \(\mathcal{F}_m\) as shown in figure~\ref{fig:2criticalWalk}(c) while increasing the same step-length will cause the walk to be in \(\mathcal{F}_j\).
Similarly, a small decrease in a length that is an upward step in \({w^*_{j}}\) will cause the walk to be in \(\mathcal{F}_j\) as shown in figure \ref{fig:2criticalWalk}(d) while increasing the same step will cause the walk leave \(\mathcal{F}_j\) and join \(\mathcal{F}_m\). Alternatively, if the walk touches zero at step \(m\) then increasing a downward length will cause it to be in \(\mathcal{F}_m\) while the paired walk will leave \(\mathcal{F}_m\).

Figure \ref{fig:2criticalWalk}(a) and (b) illustrates paired critical walks that touch the origin at step eight. When the step illustrated by the double red line is shortened, the trajectory in (c) leaves the Event of passage at step eight and \(z_8>0\) as indicated by red dot above the axis.  The paired walk in (d) enters the Event as the dot moves below the line indicating \(z_8<0\).

If a step of length \(c^*_k\) in sub-walk \(w^*_j\) has, say, a negative sign, then it is in \(w'^*_j\) with a positive sign.
When \(c^*_k\) is changed to a lower value then walk \(w'\) is in \(\mathcal{F}_m\) and \(w\) is not.
Similarly, when \(c_k\) is changed to a higher value, then walk \(w\) is in \(\mathcal{F}_m\) and \(w'\) is not.
Thus as \(c_k\) is changed and the set of lengths crosses \(\mathcal{B}_n\) one of each pair of walks leaves the Event \(\mathcal{F}_m\) and the other joins the Event.
Although the pair of walks switch which one is in Event \(\mathcal{F}_m\), the total contribution of the two walks to the cardinality of the Event is the same.
The cardinality of \(\mathcal{F}_m\) is therefore unchanged by crossing \(\mathcal{B}_n\).
Examining the complete set of pairs of walks with time-reversed sub-walks thus shows that the cardinality of first-passage Events, and thus the probability of first-passage at step \(2m\) is independent of the set of \(n\) real lengths, except for the boundary set \(\mathcal{B}_n\) of measure \(0\) in \(\mathcal{C}_n\).
Now averaging this invariant probability over all sets of lengths in \(\mathcal{C}_n\) gives the result that the distribution of first-passage step for alternating walks is step-length distribution-free.

The sub-walk must have even length for alternating walks so that walks beginning with \(w\) and \(w'\) are both alternating walks in our sample space.
First-passage always occurs on an even number step for alternating walks, so that is not a problem here.
In other types of Events on \(\mathcal{A}_n\) the requirement that the sub-walks must have even length is important.
The argument for first-passage statistics for symmetric walks carries through the same as for alternating walks with the exception that the location of the first-passage, and the length of the relevant sub-walks, need not be even.

To calculate first-passage probabilities for an alternating walk, any set of lengths suffices.
Make a convenient choice like any subset of integer powers of \(2\) where each length is larger than the sum of all smaller lengths.
First-passage of an alternating walk occurs only in a valley, i.e., on even numbered steps \(m=2m_p\).
Given a set of \(2n\) lengths selected from integer powers of \(2\), we show below that fraction of walks generated from these lengths with first-passage at step \(n\) is
\begin{eqnarray}
P^f_{n} = \frac{C_{n}}{2^{2n-1}}
\label{eq6:pf}
\end{eqnarray}
and the fraction that remains nonnegative is
\begin{eqnarray}
P^{+}_{n} = \left(2n-1\right)\frac{C_{n-1}}{2^{2n-1}}.
\label{eq6:fpcalc}
\end{eqnarray}

The calculation will proceed by induction on \(n\).
The theorem is true for \(n = 1\) because the probability of first passage in the first valley is \(\textstyle\frac{1}{2}\), as is the probability that the walk stays positive.
Either the first step upward is larger or smaller than the second step.

Suppose the theorem is true for all \(m_p \leq n\).
Consider alternating walks generated from a set \(\mathbf{c}_{2(n+1)}\) of \(2(n+1)\) lengths selected from integer powers of \(2\). Divide the alternating walks generated into a complete set of disjoint subsets where all walks in a subset have the same last two steps.
For each of these subsets, the first \(2n\) steps are all the permutation of the same set of lengths.
By induction, the fraction of these in \(\mathcal{F}_{2m_p}\) for \(m_p \le n\) is given by the theorem.
Similarly, the fraction that stay positive until the last step is given by the theorem.

First passage can occur after the peak \(n+1\) only if the walk stayed positive for the first \(n\) valleys and step \(2n\) is the largest element of the set. The probability of this is
\begin{equation}\label{eq:6app1}
  P^f_{n+1} =\frac{P^+_{n}}{2n+2}.
\end{equation}
The walk will stay positive only if it is positive for the first \(n\) valleys and it does not have a first passage at valley \(n+1\), so
\[P^+_{n+1} =p^+_{n}-P^f_{n+1}\]
in agreement with equation~(\ref{eq6:fpcalc}).
The theorem is true for \(n+1\), and so is proved by induction.
This combinatorial approach finally proves the assertion we postulated based on the results of Monte Carlo simulations.

\section*{References}

\begin{thebibliography}{99}

\bibitem{kubelka1931article}
Paul Kubelka and Franz Munk.
\newblock An article on optics of paint layers.
\newblock {\em Z. Tech. Phys}, 12, 1931.

\bibitem{yule1951penetration}
JAC Yule and WJ~Nielsen.
\newblock The penetration of light into paper and its effect on halftone
  reproduction.
\newblock In {\em Proc. TAGA}, volume~3, pages 65--76, 1951.

\bibitem{rudnick2004elements}
Joseph Rudnick and George Gaspari.
\newblock {\em Elements of the random walk: an introduction for advanced
  students and researchers}.
\newblock Cambridge University Press, 2004.

\bibitem{redner2001guide}
Sidney Redner.
\newblock {\em A guide to first-passage processes}.
\newblock Cambridge University Press, 2001.

\bibitem{spitzer_principles_1964}
Frank Spitzer.
\newblock {\em Principles of {Random} {Walk}}, volume~34 of {\em Graduate
  {Texts} in {Mathematics}}.
\newblock Springer New York, New York, NY, 1964.
\newblock DOI: 10.1007/978-1-4757-4229-9.

\bibitem{philips-invernizzi_bibliographical_2001}
Bernadette Philips-Invernizzi, Daniel Dupont, and Claude Caze.
\newblock Bibliographical review for reflectance of diffusing media.
\newblock {\em Optical Engineering}, 40(6):1082--1092, 2001.

\bibitem{schwarzschild1906equilibrium}
K~Schwarzschild.
\newblock On the equilibrium of the sun's atmosphere.
\newblock {\em Nachrichten von der K{\"o}niglichen Gesellschaft der
  Wissenschaften zu G{\"o}ttingen. Math.-phys. Klasse, 195, p. 41-53},
  195:41--53, 1906.

\bibitem{chandrasekhar1950radiative}
Subrahmanyan Chandrasekhar.
\newblock {\em Radiative transfer.}
\newblock Oxford, Clarendon Press, 1950.

\bibitem{chandrasekhar1943stochastic}
Subrahmanyan Chandrasekhar.
\newblock Stochastic problems in physics and astronomy.
\newblock {\em Reviews of modern physics}, 15(1):1, 1943.

\bibitem{bonner1987model}
RF~Bonner, R~Nossal, S~Havlin, and GH~Weiss.
\newblock Model for photon migration in turbid biological media.
\newblock {\em JOSA A}, 4(3):423--432, 1987.

\bibitem{nossal1988photon}
Ralph Nossal, J~Kiefer, George~H Weiss, Robert Bonner, Haim Taitelbaum, and
  Sholmo Havlin.
\newblock Photon migration in layered media.
\newblock {\em Applied optics}, 27(16):3382--3391, 1988.

\bibitem{schuster1905radiation}
Arthur Schuster.
\newblock Radiation through a foggy atmosphere.
\newblock {\em The astrophysical journal}, 21:1, 1905.

\bibitem{gate1974comparison}
LF~Gate.
\newblock Comparison of the photon diffusion model and kubelka-munk equation
  with the exact solution of the radiative transport equation.
\newblock {\em Applied optics}, 13(2):236--238, 1974.

\bibitem{sandoval_deriving_2014}
Christopher Sandoval and Arnold~D. Kim.
\newblock Deriving kubelka-munk theory from radiative transport.
\newblock {\em Journal of the Optical Society of America A}, 31(3):628, 2014.

\bibitem{youngquist1987optical}
Robert~C Youngquist, Sally Carr and David Davies.
\newblock Optical coherence-domain reflectometry: a new optical evaluation
  technique.
\newblock {\em Optics letters}, 12(3):158--160, 1987.

\bibitem{haney2007modified}
Matthew~M Haney and Kasper van Wijk.
\newblock Modified kubelka-munk equations for localized waves inside a layered
  medium.
\newblock {\em Physical Review E}, 75(3):036601, 2007.

\bibitem{hebert2008correspondence}
Mathieu H{\'e}bert and Jean-Marie Becker.
\newblock Correspondence between continuous and discrete two-flux models for
  reflectance and transmittance of diffusing layers.
\newblock {\em Journal of Optics A: Pure and Applied Optics}, 10(3):035006,
  2008.

\bibitem{ballestra2016very}
Luca~Vincenzo Ballestra, Graziella Pacelli, and Davide Radi.
\newblock A very efficient approach to compute the first-passage probability
  density function in a time-changed brownian model: Applications in finance.
\newblock {\em Physica A: Statistical Mechanics and its Applications},
  463:330--344, 2016.

\bibitem{simon2003random}
Klaus Simon and Beat Trachsler.
\newblock A {{Random Walk Approach}} for {{Light Scattering}} in {{Material}}.
\newblock {\em Discrete Mathematics and Theoretical Computer Science}, pages
  289--300, 2003.

\bibitem{jacques_monte_2010}
Steven~L. Jacques.
\newblock Monte {{Carlo}} modeling of light transport in tissue (steady state
  and time of flight).
\newblock In {\em Optical-Thermal Response of Laser-Irradiated Tissue}, pages
  109--144. {Springer}, 2010.

\bibitem{doering1992long}
Charles~R Doering, Tane~S Ray, and M~Lawrence Glasser.
\newblock Long transmission times for transport through a weakly scattering
  slab.
\newblock {\em Physical Review A}, 45(2):825--828, 1992.

\bibitem{antal2006escape}
T~Antal and S~Redner.
\newblock Escape of a uniform random walk from an interval.
\newblock {\em Journal of statistical physics}, 123(6):1129--1144, 2006.

\bibitem{wuttke2014zig}
Joachim Wuttke.
\newblock The zig-zag walk with scattering and absorption on the real half line
  and in a lattice model.
\newblock {\em Journal of Physics A: Mathematical and Theoretical},
  47(21):215203, 2014.

\bibitem{darwin1922xcii}
CG~Darwin.
\newblock Xcii. the reflexion of x-rays from imperfect crystals.
\newblock {\em The London, Edinburgh, and Dublin Philosophical Magazine and
  Journal of Science}, 43(257):800--829, 1922.

\bibitem{hamilton1957effect}
Walter~C Hamilton.
\newblock The effect of crystal shape and setting on secondary extinction.
\newblock {\em Acta Crystallographica}, 10(10):629--634, 1957.

\bibitem{andersen1962equivalence}
E~Sparre Andersen.
\newblock The equivalence principle in the theory of fluctuations of sums of
  random variables.
\newblock In {\em Colloquium on Combinatorial Methods in Probability Theory,
  Aarhus}, pages 13--16, 1962.

\bibitem{AndersenFluctuations}
E.~Sparre~Andersen.
\newblock On the fluctuations of sums of random variables.
\newblock {\em Mathematica Scandinavica}, 1:263--285, 1953.

\bibitem{myrick2011kubelka}
Michael~L Myrick, Michael~N Simcock, Megan Baranowski, Heather Brooke,
  Stephen~L Morgan, and Jessica~N McCutcheon.
\newblock The kubelka-munk diffuse reflectance formula revisited.
\newblock {\em Applied Spectroscopy Reviews}, 46(2):140--165, 2011.

\bibitem{beer1852bestimmung}
August Beer.
\newblock Bestimmung der absorption des rothen lichts in farbigen
  flussigkeiten.
\newblock {\em Ann. Physik}, 162:78--88, 1852.

\bibitem{lambert1760photometria}
Johann~Heinrich Lambert.
\newblock Photometria, sive de mensura et gradibus luminis.
\newblock {\em Colorum et Umbrae, Eberhard Klett, Augsberg, Germany}, 1760.

\bibitem{OEIS2019}
et~al N~J A~Sloane.
\newblock The on-line encyclopedia of integer sequences.

\end{thebibliography}
\bibliographystyle{iopart-num}

\end{document}